# MACH'S PRINCIPLE AND HIGHER-DIMENSIONAL DYNAMICS

B. Mashhoon[1] and P.S. Wesson[2,3]

[1]Department of Physics and Astronomy, University of Missouri, Columbia, Missouri 65211, USA

[2]Department of Physics and Astronomy, University of Waterloo, Waterloo, Ontario N2L 3G1, Canada

[3]Herzberg Institute of Astrophysics, National Research Council, Victoria, B.C. V9E 2E7, Canada

Abstract

We briefly discuss the current status of Mach's principle in general relativity and point out that its last vestige, namely, the gravitomagnetic field associated with rotation, has recently been measured for the earth in the GP-B experiment. Furthermore, in his analysis of the foundations of Newtonian mechanics, Mach provided an operational definition for inertial mass and pointed out that time and space are conceptually distinct from their operational definitions by means of masses. Mach recognized that this circumstance is due to the lack of any a priori connection between the inertial mass of a body and its Newtonian state in space and time. One possible way to improve upon this situation in classical physics is to associate mass with an extra dimension. Indeed, Einstein's theory of gravitation can be locally embedded in a Ricci-flat 5D manifold such that the 4D energy-momentum tensor appears to originate from the existence of the extra dimension. An outline of such a 5D Machian extension of Einstein's general relativity is presented.

I. Introduction

In his treatise on the science of mechanics, Ernst Mach provided a critical analysis of the physical foundations of Newtonian mechanics [1]. The epistemological difficulties associated with the Newtonian notions of absolute space and time, and the ensuing problem of the physical origin of inertial forces, led Mach to suggest that inertia is due to mutual interaction between masses. Mach's ideas provided the early inspiration for Einstein on his path to general relativity. It is now generally acknowledged, however, that what may loosely be termed Mach's principle is not properly incorporated into general relativity; in particular, the origin of inertia remains essentially the same as in Newtonian physics. Various attempts at the resolution of difficulties that are encountered in linking Mach's principle with Einstein's theory of gravitation have led to many interesting investigations [2-6]; moreover, a perusal of these references illustrates the diversity of opinion regarding Mach's principle that still lingers.

Within the framework of Einstein's theory of gravitation, the current status of Mach's principle, especially in connection with the physical origin of inertial forces, is discussed in sections II and III. Briefly, in Einstein's theory, the global inertial frames of Newtonian physics are replaced by local inertial frames in accordance with Einstein's *local* principle of equivalence; nevertheless, so long as only Newtonian gravitational accelerations—due to their universality—can be rendered "relative" in general relativity, non-gravitational accelerated motion retains its absolute character in Einstein's theory of gravitation. Classical inertial effects thus originate from accelerated motion with respect to the local inertial frames.

It is important to recognize that in the course of his critique of Newton's view of time, space and motion, Mach also identified the essential epistemological weakness of Newtonian mechanics: The *internal* state of a Newtonian point particle, characterized by its inertial mass, has no immediate connection with the particle's *external* state in absolute space and time, characterized by its position and velocity [1]. As described in detail in section IV below, Mach considered how time and space are *operationally* defined via masses and concluded that time and space, as physical concepts, are in fact different from their operational definitions by means of masses. That is, in Newtonian mechanics, masses are simply "placed" in absolute space and time, which remain external to them. As a consequence of this circumstance, the extrinsic state of a point mass *m*, namely its position **x** and velocity **v**, can be shared by other "comoving" masses.

How can this lack of organic connection between *m* and (**x**, **v**) be overcome, apart from Mach's own solution in terms of relativity of all motion? One way involves the standard extension of classical mechanics to the quantum domain; indeed, this line of thought entails complementarity of absolute and relative motion and has been discussed in references [7-12]. Another way is to associate the internal states of classical particles with an extra dimension. That is, Einstein's theory of gravitation can be extended to a 5D space-time-matter theory such that the vacuum gravitational field in 5D appears in 4D with an induced energy-momentum tensor as the source of the 4D gravitational field. Indeed, every solution of Einstein's theory in 4D can be locally embedded in a Ricci-flat 5D manifold in accordance with Campbell's theorem [13-17]. In such a theory, the free motion of a classical particle in 5D would in general involve the variation of space, time and inertial mass along its 4D world line; therefore, the inertial mass of a test particle varies in space-time. This is in contrast to standard general relativity, where the mass of a test particle is independent of space-time [18, 19]. We briefly describe such a 5D Machian extension of general relativity in section V.

II. Relativity of Accelerated Motion?

Ernst Mach had a predilection for *relativity*. In a collection of three essays that he wrote on space and geometry, we find in his discussion of "the relativity of all spatial relations" [20] that:

"…All *physical* determinations are *relative*. Consequently, likewise all *geometrical* determinations possess validity only *relatively* to the measure. The concept of measurement is a concept of relation, which contains nothing not contained in the measure…."

It is therefore clear that the absolute space and time of Newtonian physics posed a dilemma for Mach: it was impossible for him to reconcile the great practical success of Newtonian mechanics with the fact that it was based on the notions of absolute space and absolute time, which Mach considered to be unphysical. He imagined that these only formed a temporary framework for physics and would be replaced someday by a future theory based purely on the relativity of all motion [1].

In Newtonian mechanics, absolute motion is defined as movement with respect to the ensemble of inertial frames of reference that are related to each other by Galilean transformations. Moreover, the reality of absolute space and time is revealed via the existence of *inertial forces*, which are experienced, for instance, by observers at rest in reference systems that accelerate with respect to inertial frames of reference. Mach speculated that the acceleration must instead be *relative* to the rest of the matter in the universe [1]. That is, inertial forces must be due to an interaction between local and distant masses. Following Mach, Einstein suggested that inertial forces could be of gravitational origin. Such a *global* Machian equivalence between inertia and gravitation would be consistent with the principle of equivalence of inertial and gravitational masses, since inertial forces acting on a body are all proportional to its inertial mass and hence, via the equivalence principle, its gravitational mass as well. To explain the motivation for his pioneering 1918 work on the gravitational effects of rotating masses in general relativity, Hans Thirring quoted from a 1914 paper of Einstein as follows [21]:

"At first it seems that such an extension of the theory of relativity has to be rejected for physical reasons. Namely, let *K* be a permissible coordinate system in the Galilei-Newtonian sense, *K'* a uniformly rotating coordinate system with respect to *K*. Then centrifugal forces act on masses which are at rest relative to *K'*, while not on masses at rest relative to *K*. Already Newton considered this as proof that the rotation of *K'* has to be interpreted as "absolute", and that one cannot thus treat *K'* as "at rest" with the same justification as *K*. This argument is not valid, however, as explained by E. Mach in particular. That is, we need not attribute the existence of those centrifugal forces to the motion of *K'*; rather, we can attribute them as well to the average rotational motion of the ponderable distant masses in the surrounding relative to *K'*, whereby we treat *K'* as at rest. If the Newtonian laws of mechanics do not admit such a conception, this could well be caused by the defects of this theory…."

In short, relativity of accelerated motion necessitates a new theory of gravitational interactions. On the other hand, if such a theory is formulated as a relativistic field theory, then there is an insurmountable conflict with causality as the retardation times between local events and distant masses would be extremely long. Even in our galaxy, it would take thousands of years for local news to reach the stars of our galaxy and similarly for their reaction to get back to earth.

Einstein's theory of gravitation, which is in good agreement with solar system data, is indeed a relativistic field theory modeled after Maxwell's theory of the electromagnetic field. The modern generally accepted geometric formulation of Einstein's theory is based upon the *local* equivalence of an observer in a gravitational field with a certain accelerated observer in Minkowski space-time—in fact, this is the content of Einstein's heuristic principle of equivalence. Motivated in its infancy by the Machian notion of relativity, the current theory of gravitation is Einstein's general relativity in which the gravitational field is identified with the Riemannian curvature of space-time [22].

III. Mach's Principle

To save the essence of Mach's idea within the context of general relativity theory, we can conceive of two different universes, one would be our universe in which a system $S$ accelerates and the expected inertial forces immediately appear in $S$ and another one in which system $S$ does not accelerate but instead the rest of the matter in the universe accelerates in the opposite direction in just such a way as to produce the same relative motion. The question is then whether in the latter universe, after a sufficiently long time, gravitational forces appear in $S$ that are the same as the standard inertial forces in our universe. This is the gist of the formulation of Mach's principle considered by Thirring and subsequent investigators [21, 23]; it is certainly much weaker than Mach's notion of relativity of all motion, but opens the way to the theoretical investigation of Machian effects in general relativity.

Consider, for instance, a simple situation in which $S$ linearly accelerates along the negative $z$-axis of an inertial reference frame and the attendant inertial forces then act on masses in $S$ along the positive $z$-axis. Now if in the other universe all the distant masses accelerate along the $z$ direction, we expect on the basis of Mach's principle that gravitational forces would eventually appear in $S$ and act along the same direction. In this way, a simple consequence of the law of inertia in the standard Newtonian analysis acquires, in the new interpretation, the status of a Machian induction effect.

If inertia is due to interaction between masses, then one would expect, in accordance with Einstein's formulation of Mach's principle [24], that the inertial mass of a particle would increase with proximity to other masses. Einstein expected that such an increase in the inertial mass, even if very small, would show up in general relativity [24]. Indeed, we find on page 100 of Einstein's book [24] the following three Machian expectations:

"…What is to be expected along the line of Mach's thought?

1. The inertia of a body must increase when ponderable masses are piled up in its neighbourhood.

2. A body must experience an accelerating force when neighbouring masses are accelerated, and, in fact, the force must be in the same direction as the acceleration.

3. A rotating hollow body must generate inside of itself a "Coriolis field," which deflects moving bodies in the sense of the rotation, and a radial centrifugal field as well."

Brans has shown that the inertial mass of a free test particle cannot change in a gravitational field if one adheres to the modern geometric interpretation of Einstein's theory of gravitation. That is, no extra inertia is induced in a body as a result of the presence of other bodies. In fact, Brans's thorough analysis has completely settled this issue [18, 19]. The situation is somewhat different with respect to the other two Machian predictions: quantitative predictions based on general relativity are in general more properly interpreted in terms of tidal dynamics and gravitomagnetism without any recourse to Mach's principle. To clarify some of the issues involved here, let us next consider the second point, which involves another aspect of Machian inertial induction and is also related to the question of the origin of inertial forces. Einstein's field theory of gravitation makes it possible to describe in general terms the relative motion of local masses—via generalizations of the Jacobi equation—due to the gravitational field of accelerating neighboring masses. The resulting tidal effects do *not* in general coincide with Machian inertial induction effects. Moreover, *gravitational induction* can be theoretically explored on the basis of general relativity in close analogy with Faraday's law of induction in electrodynamics [25]. The important point here is that, even in this weaker formulation, inertial forces are not of gravitational origin within the generally accepted framework of Einstein's theory of gravitation. Indeed, Einstein's local principle of equivalence ensures that all observers have access to local inertial frames, where the laws of non-gravitational physics are locally valid. Though the global inertial frames of Newtonian physics are now replaced by local inertial frames, non-gravitational accelerated motion retains its absolute character in general relativity.

Basically, what remains of the above three points is the "Coriolis field," which is essentially the gravitomagnetic field generated by the *absolute* rotation of a body. This field is also responsible for the "dragging of the local inertial frames." Indeed, the exterior gravitomagnetic field of the earth has been recently measured via Gravity Probe B (GP-B) experiment [26]. Einstein's local principle of equivalence, within the general *linear* approximation of general relativity, is equivalent to the gravitational Larmor theorem [27], so that the gravitomagnetic field can be locally replaced by a rotation. Hence the gravitomagnetic field is locally equivalent to a "Coriolis field."

IV. Mach's Critique of Newtonian Physics

Mach was a pioneer in novel methods of teaching physics and a strong proponent of the use of thought experiments as well as operational definitions of physical quantities [28]. Indeed, in classical mechanics, Mach provided the operational definition of inertial mass via Newton's third law of motion. Mach also considered the operational definition of inertial frames of reference, which represent absolute space and time and are essential in the formulation of Newtonian mechanics.

In commenting upon the efforts of various investigators to make sense of the law of inertia by constructing inertial frames of reference either via thought experiments or empirically via astronomical observations, Mach noted the fundamental difficulty associated with the operational definitions of time and space in terms of masses on pages 295-296 of [1] as follows:

"…Although I expect that astronomical observation will only as yet necessitate very small corrections, I consider it possible that the law of inertia in its simple Newtonian form has only, for us human beings, a meaning which depends on space and time. Allow me to make a more general remark. We measure time by the angle of rotation of the earth, but could measure it just as well by the angle of rotation of any other planet. But, on that account, we would not believe that the *temporal* course of all physical phenomena would have to be disturbed if the earth or the distant planet referred to should suddenly experience an abrupt variation of angular velocity. We consider the dependence as not immediate, and consequently the temporal orientation as *external*. Nobody would believe that the chance disturbance—say by an impact—of one body in a system of uninfluenced bodies which are left to themselves and move uniformly in a straight line, where all the bodies combine to fix the system of coördinates, will immediately cause a disturbance of the others as a consequence. The orientation is external here also. Although we must be very thankful for this, especially when it is purified from meaninglessness, still the natural investigator must feel the need of further insight—of knowledge of the *immediate* connections, say, of the masses of the universe. There will hover before him as an ideal an insight into the principles of the whole matter, from which accelerated and inertial motions result in the *same* way. The progress from Kepler's discovery to Newton's law of gravitation, and the impetus given by this to the finding of a physical understanding of the attraction in the manner in which electrical actions at a distance have been treated, may here serve as a model. We must even give rein to the thought that the masses which we see, and by which we by chance orientate ourselves, are perhaps not those which are really decisive…."

It follows from Mach's considerations that the physical concepts of time and space are distinct from their operational definitions by means of masses. This is because there is no a priori connection between a particle's mass and its representation in time and space. Mach's account of these issues brings to mind, among other things, modern time determination via pulsar timing and position determination via the Global Positioning System [29].

Quantum mechanics is a generalization of classical mechanics, and it has been argued that the epistemological problem of classical physics elucidated by Mach is ameliorated in the quantum theory [8-12]. On the other hand, to overcome the difficulty within *classical* physics, it appears necessary to associate, in a general sense, mass with an extra dimension. That is, mass may be married to the space-time coordinates by extending the dimensionality of gravitational theory. This involves a special Kaluza-Klein extension of Einstein's theory of gravitation from 4D to 5D that generally results in the variation of particle masses in space-time [30-32]. It is briefly outlined in the following section in connection with Mach's critique of Newtonian dynamics.

V. Space-Time-Matter

An event can be uniquely characterized by its space-time coordinates $x^\alpha = (ct,x,y,z)$. Here Greek indices run from 0 to 3 and henceforth we set $c$, the speed of light in vacuum, equal to unity, unless otherwise specified. We consider a 5D Riemannian extension of space-time such that the 5D metric has the general form

$$dS^2 = {}^{(5)}g_{AB} dx^A dx^B, \qquad A,B = 0, 123, 4. \tag{1}$$

The freedom in the choice of coordinates may be employed to set ${}^{(5)}g_{44} = -1$ and ${}^{(5)}g_{\alpha 4} = 0$ via a procedure that is entirely analogous to the construction of the synchronous reference system in 4D [33]. We thus arrive at the standard canonical metric of 5D relativity

$$dS^2 = \frac{\ell^2}{L^2} g_{\alpha\beta}(x^\gamma, \ell) dx^\alpha dx^\beta - d\ell^2, \tag{2}$$

where $x^4 = \ell$ and $L$ is a constant length that has been introduced on the basis of dimensional considerations. The 4D space-time interval is then *defined* by

$$ds^2 = g_{\alpha\beta}(x^\gamma, \ell) dx^\alpha dx^\beta, \tag{3}$$

so that $dS^2 = (\ell/L)^2 ds^2 - d\ell^2$. The signature of the space-time metric is -2 in our convention.

There are no material sources in 5D; therefore, the 5D gravitational field equations reduce to ${}^{(5)}R_{AB} = 0$. The components of the 5D Ricci tensor for the canonical metric (2) are

$$^{(5)}R_{44} = -\frac{\partial A^\alpha{}_\alpha}{\partial \ell} - \frac{2}{\ell} A^\alpha{}_\alpha - A_{\alpha\beta} A^{\alpha\beta}, \tag{4}$$

$$^{(5)}R_{\mu 4} = A_\mu{}^\alpha{}_{;\alpha} - \frac{\partial \Gamma^\alpha_{\mu\alpha}}{\partial \ell}, \tag{5}$$

$$^{(5)}R_{\mu\nu} = R_{\mu\nu} - S_{\mu\nu}, \tag{6}$$

where $A^\alpha{}_\beta = g^{\alpha\gamma} A_{\gamma\beta}$ and $A_{\alpha\beta}$ is a symmetric space-time tensor defined by

$$A_{\alpha\beta} = \tfrac{1}{2} \frac{\partial g_{\alpha\beta}}{\partial \ell}. \tag{7}$$

Furthermore, a semicolon represents the usual 4D covariant derivative, $R_{\mu\nu}$ and $\Gamma^{\mu}_{\nu\rho}$ are respectively the 4D Ricci tensor and Christoffel symbols determined from $g_{\mu\nu}(x^{\rho},\ell)$ and

$$S_{\mu\nu} = \frac{\ell^2}{L^2}\left[\frac{\partial A_{\mu\nu}}{\partial \ell} + \left(\frac{4}{\ell} + A^{\alpha}{}_{\alpha}\right)A_{\mu\nu} - 2A_{\mu}{}^{\alpha}A_{\nu\alpha}\right] + \frac{1}{L^2}(3 + \ell A^{\alpha}{}_{\alpha})g_{\mu\nu}. \tag{8}$$

The ten metric functions of five variables, $g_{\alpha\beta}(x^{\gamma},\ell)$, satisfy the fifteen partial differential equations that are obtained from setting the right-hand sides of Eqs. (4)-(6) equal to zero. Moreover, the effective 4D energy-momentum tensor that is induced by the extra spacelike dimension can be obtained from Eqs. (6) and (8) via Einstein's field equations, namely, $T_{\alpha\beta} = G_{\alpha\beta}/(8\pi G)$, where $G_{\alpha\beta}$ is the corresponding 4D Einstein tensor.

Let us now consider the motion of a free test particle in 5D. The 5D geodesic equation can be expressed in terms of 4D equation of motion

$$\frac{du^{\mu}}{ds} + \Gamma^{\mu}_{\alpha\beta}u^{\alpha}u^{\beta} = (-g^{\mu\alpha} + \tfrac{1}{2}u^{\mu}u^{\alpha})u^{\beta}\frac{\partial g_{\alpha\beta}}{\partial \ell}\frac{d\ell}{ds} \tag{9}$$

and the variation of the fifth coordinate $\ell$ along the world line

$$\frac{d^2\ell}{ds^2} - \frac{2}{\ell}\left(\frac{d\ell}{ds}\right)^2 + \frac{\ell}{L^2} = -\tfrac{1}{2}\left[\frac{\ell^2}{L^2} - \left(\frac{d\ell}{ds}\right)^2\right]u^{\alpha}u^{\beta}\frac{\partial g_{\alpha\beta}}{\partial \ell}. \tag{10}$$

Here $u^{\alpha} = dx^{\alpha}/ds$ is the particle's unit 4-velocity in space-time; in fact,

$$g_{\alpha\beta}(x^{\gamma},\ell)u^{\alpha}u^{\beta} = 1. \tag{11}$$

This relation turns out to be consistent with Eq. (9).

The absence of any material source in 5D implies the natural assumption that the free test particle follows a null path in 5D. Indeed, a 5D null ray can correspond to a massive particle following a timelike world line in 4D. It follows from $dS^2 = 0$ and Eq. (2) that $(d\ell/ds)^2 = \ell^2/L^2$. Any solution of this equation, namely, $\ell = \ell_0 \exp(\pm s/L)$, where $\ell_0$ is a constant, renders both sides of Eq. (10) equal to zero. Hence for a null geodesic in 5D, Eq. (10) is automatically satisfied. The null interval in 5D corresponds to conventional causality in 4D. For photons, for instance, $\ell_0 = 0$, so that their 4D null paths are confined to hypersurface $x^4 = \ell = 0$.

We now turn our attention to the physical content of Eq. (9) for a massive test particle. The canonical coordinates under consideration in 5D are such that there is complete freedom in the choice of space-time coordinates, just as in Einstein's general relativity. Hence, as is well known [33], it is always possible to introduce locally geodesic coordinates near an event in 4D space-time such that the Christoffel symbols, $\Gamma^{\mu}_{\alpha\beta}(x^{\gamma},\ell)$, all vanish at the event and the space-time metric tensor, $g_{\alpha\beta}(x^{\gamma},\ell)$, assumes its Minkowski form, namely, diag(1, -1, -1, -1). In this way, an observer in 4D space-time, where observational data are analyzed, can always set up a *local inertial frame* about an event in space-time. It follows from these considerations, where $\ell$ is simply treated as a parameter, that in any such local inertial frame, Eq. (9) may be compared with the special relativistic form of Newton's second law of motion. In special relativity, the general force law for the motion of a massive test particle in an inertial frame can be written as

$$\frac{dp^{\mu}}{ds} = F^{\mu}, \tag{12}$$

where $p^{\mu} = mu^{\mu}$ is the particle's 4-momentum, $m$ is its inertial mass and $F^{\mu}$, $F^{\mu}u_{\mu} = 0$, is the net 4-force acting on the particle [33]. In the particle's rest frame, all basic forces are 3D vectors, hence the 4-force must be orthogonal to the particle's 4-velocity. Comparing Eq. (9), expressed in a local inertial frame, with Eq. (12), we then find that in arbitrary space-time coordinates

$$F^{\mu} = m(-g^{\mu\alpha} + u^{\mu}u^{\alpha})u^{\beta}\frac{\partial g_{\alpha\beta}}{\partial \ell}\frac{d\ell}{ds}, \tag{13}$$

$$\frac{1}{m}\frac{dm}{ds} = \tfrac{1}{2}u^{\alpha}u^{\beta}\frac{\partial g_{\alpha\beta}}{\partial \ell}\frac{d\ell}{ds}. \tag{14}$$

The existence of an extra force $F^{\mu}$ of gravitational origin due to the dependence of the space-time metric on the fifth coordinate is a clear violation of Einstein's principle of equivalence in our 4D space-time. Moreover, it follows from Eq. (14) that the inertial mass of the particle varies along its world line, that is, in general $m = m(t, \mathbf{x})$. Relations (13) and (14) represent new aspects of the gravitational interaction in this framework [34-36].

To illustrate these results, consider the special case in which the space-time metric is independent of the fifth coordinate, that is, $g_{\alpha\beta} = g_{\alpha\beta}(x^{\gamma})$. Then, $A_{\alpha\beta} = 0$ in the field equations, so that the 5D Ricci-flat requirement simply reduces to $R_{\mu\nu} = (3/L^2)g_{\mu\nu}$. This means that the 4D space-time is an Einstein space with cosmological constant $\Lambda = 3/L^2$. It follows that any vacuum solution of Einstein's theory with a nonzero cosmological constant, such as the Kerr-de Sitter solution, automatically generates a canonical solution of the 5D field equations with $L = (3/\Lambda)^{1/2}$. Furthermore, the equations of motion for a test particle in the 4D part of such a 5D manifold are identical to those in general relativity. That is, $F^{\mu} = 0$ and the inertial mass is a constant along the 4D geodesic. This means that the classical tests of relativity are satisfied in this case. Let us note that the equation of motion along the extra axis of the 5D metric, $\ell = \ell_0 \exp(\pm s/L)$, implies that a massive test particle wanders away from a given $\ell$-hypersurface at a slow rate governed by the cosmological constant.

In general, $\partial g_{\alpha\beta}/\partial \ell \neq 0$, and it is then a general property of the 5D theories under consideration that the inertial mass of a massive test particle changes over cosmological time. The key idea, originally developed in [37-39], which has been the starting point of this particular approach to Kaluza-Klein theory, is the association of matter with the extra dimension. The general expression for the 4D energy-momentum tensor for all solutions of the 5D Ricci-flat field equations was first derived in 1992 [39]. The various implications of this theory have been presented in detail in recent monographs [30, 31].

While canonical coordinates are convenient in that they correspond to the traditional way of doing dynamics, it should be recalled that in principle any choice of coordinates is acceptable in a covariant 5D theory, and for most of these the physics may not be readily identifiable. In particular, we note that the canonical coordinate conditions do not completely fix the canonical coordinate system. That is, the space-time coordinates can be transformed among themselves and the fifth coordinate is determined only up to a simple translation. The latter is analogous to the arbitrariness in the choice of the origin of time in the synchronous coordinates of Einstein's general relativity. In 5D relativity, replacing $\ell$ by $\ell - a$, where $a$ is a constant parameter, in the canonical metric (2) has significant physical implications that have been recently explored [40-42].

Finally, let us consider the possibility that the inertial mass expressed as a length may in fact be chosen as the fifth coordinate. This is indeed permitted by general covariance once it becomes clear via Eq. (14) that mass in general varies with the fifth coordinate. To this end, let

$$\lambda = \frac{Gm}{c^2} \qquad (15)$$

be the gravitational length that can be chosen as the fifth coordinate such that 4D physics is recovered on the hypersurface $x^4 = \lambda =$ constant. The motion of a free test particle in 4D general relativity is given by the variational principle of stationary action, namely,

$$\delta \int -mc\, ds = 0, \qquad (16)$$

where $ds$ is the standard 4D space-time interval. One way to incorporate into this variational principle the idea of a variable mass as the fifth dimension would be to extend Eq. (16), up to constant factors, as follows:

$$\delta \int \lambda \left[ g_{\alpha\beta}(x^\gamma, \lambda) dx^\alpha dx^\beta - h(x^\gamma, \lambda) d\lambda^2 \right]^{\frac{1}{2}} = 0, \qquad (17)$$

where $h$ is a new dimensionless metric function. Comparing this relation with the variational principle for geodesic motion in 5D, namely,

$$\delta \int dS = 0, \qquad (18)$$

naturally leads to

$$dS^2 = \frac{\lambda^2}{L^2} g_{\alpha\beta}(x^\gamma, \lambda) dx^\alpha dx^\beta - \Phi(x^\gamma, \lambda) d\lambda^2, \qquad (19)$$

where $L$ is a (cosmic) length whose introduction is necessary on dimensional grounds and

$$\Phi(x^\gamma, \lambda) = \frac{\lambda^2}{L^2} h(x^\gamma, \lambda) \qquad (20)$$

is a scalar field in 4D space-time. For $\Phi = 1$, Eq. (19) reduces to the canonical metric (2) with $\lambda = \ell$. The physical consequences of the assumption that 5D Riemannian metric (19) satisfies $^{(5)}R_{AB} = 0$ have been extensively studied [30, 31]. This line of thought constitutes a parallel approach to 5D space-time-matter theory [29, 43, 44].

The notion that matter exists in 4D as a consequence of source-free dynamics in 5D therefore results in a Machian version of Kaluza-Klein theory, since it overcomes Mach's fundamental epistemological objection to Newtonian physics. In this theory, the inertial mass of a test particle in general varies along its 4D world line [45].

VI. Discussion

We have presented the main aspects of a Machian approach to Kaluza-Klein theory in which matter in 4D is due to the existence of a fifth dimension and consequently the inertial mass of a test particle in general varies in space and time. This relativistic theory of gravitation takes a simple form in a canonical coordinate system that we have adopted throughout this paper. Our treatment is based on the idea that particles travel on null paths of a 5D space-time-matter manifold that is described by source-free field equations such that matter and dynamics in 4D space-time are induced by the higher geometry. This theory is mathematically somewhat similar to the more recent membrane theory, in which matter is concentrated on a singular hypersurface that is identified with space-time [46]. Space-time-matter theory was originally proposed as a way of unifying the gravitational field with its material source in a classical framework; however, the theory may have further implications in the quantum domain [47].